\documentstyle{article}[12pt]
\topmargin 10pt \textheight 225mm \textwidth 160mm \vspace{1.8cm}
\begin{document}

\baselineskip=24pt

\title{The new wave equations of relativistic and non-relativistic quantum mechanics}
\author{Xiangyao Wu$^{1}$, Yiqing Guo$^{1}$,
\thanks{The E-mailing address:
guoyq@mail.ihep.ac.cn}, Xinguo Yin$^{1}$ \\
{\footnotesize 1. Instituten of Physics, Chinese University,
Beijing 100065, China} }

\date{}
\maketitle

\renewcommand{\thesection}{Sec. \Roman{section}} \topmargin 10pt
\renewcommand{\thesubsection}{ \arabic{subsection}} \topmargin 10pt
{\vskip 5mm
\begin {minipage}{140mm}
\centerline {\bf Abstract} \vskip 8pt
\par
\indent In this work, we give the wave equations of relativistic
and non-relativistic quantum mechanics which are different from
the Schr\"{o}dinger and Klein-Gordon equation, and we also give
the new relativistic wave equation of a charged particle in an
electromagnetic field.
\end {minipage} }

\vspace*{2cm} {\bf PACS numbers: 03.00.00, 03.65.Pm, 03.65.Ta}

\newpage
We know De Broglie suggested that not only does light have a dual
nature but material particles also require a wave-particle
description during 1922-23, and he further noticed correspondences
between the classical theory of light and the classical theory of
mechanics. He thought that we may obtain the wave equation of
material particles by comparing the classical theory of light with
the classical theory of material particle \cite{s1}.
Schr\"{o}dinger used De Broglie's idea to obtain the wave equation
of material particles, i.e., Schr\"{o}dinger equation \cite{s2}.
In the following, we also apply De Broglie's suggestion to
research the wave equation of material particles \cite{s3}.

From the classical electromagnetic theory, the wave equation of
electromagnetic wave which frequency is $\nu$ is:
\begin{equation}
\nabla^2 \psi(\vec{r})+\frac{4\pi^2n^2\nu^2}{c^2}\psi(\vec{r})=0,
\end{equation}
where $n$ is refracting power, $c$ is light velocity and $\psi(r)$
is a component of electromagnetic field $E$ and $B$. The Eq. (1)
describes the wave nature of light such as interference and
diffraction phenomenon of light. When light transmits at straight
line it is described by the geometrical optics and the geometrical
optics is a limiting case of wave theory of light. Fermat had
reduced the laws of geometrical optics to the principles of
'least-time'. That is, a light ray follows the path requiring the
least time. The Fermat principle is
\begin{equation}
\delta \int n ds=0,
\end{equation}
For a material particle, when it moves in potential energy $V(r)$
it can be described by the minimum action principle of Jacobi
\begin{equation}
\delta \int \sqrt{2m(E-V(r))}ds=0,
\end{equation}
where the total energy $E$ is the sum of kinetic $T$ and potential
energy $V(r)$. When $n$ is replaced with $\sqrt{2m(E-V(r))}$, the
Eq. (2) is the same as Eq. (3). So, we can think that material
particle wave equation is similar to Eq. (1) and the wave equation
can be written as follows:
\begin{equation}
\nabla^2 \psi(\vec{r})+Cm(E-V(r))\psi(\vec{r})=0,
\end{equation}
where $C$ is a constant and it can be obtained in the following.
For a free material particle, its potential energy $V(r)=0$ and
total energy $E=\frac{p^2}{2m}$, and it is associated with a plane
wave
\begin{equation}
\psi(\vec{r})=\frac{1}{(2\pi \hbar)^{3/2}}
e^{\frac{i}{\hbar}(\vec{p}\cdot\vec{r}-Et)},
\end{equation}
Substitution of Eq. (5) into Eq. (4) gives
\begin{equation}
(\frac{i}{\hbar}p)^2\psi(\vec{r})+CmE\psi(\vec{r})=0,
\end{equation}
The constant $C$ is
\begin{equation}
C=2/{\hbar}^2,
\end{equation}
From Eq. (4) and Eq. (7), we obtain
\begin{equation}
[-\frac{{\hbar}^2}{2m}\nabla^2+V(r)]\psi(\vec{r})=E\psi(\vec{r}).
\end{equation}
The Eq. (8) is known as Schr\"{o}dinger's time-independent wave
equation and gives the states of constant energy \cite{s2}. It is
corresponding Eq. (1) which describes the electromagnetic wave of
constant energy. The time-dependent wave equation of material
particle can be obtained by time-dependent wave equation of
electromagnetic wave. It is
\begin{equation}
\nabla^2
\psi(\vec{r},t)-\frac{n^2}{c^2}\frac{\partial^2\psi(\vec{r},t)}{\partial
t^{2}}=0,
\end{equation}
As the same process in the above, we can write down the
time-dependent wave equation of material particle
\begin{equation}
\nabla^2
\psi(\vec{r},t)+Cm(E-V(r))\frac{\partial^2\psi(\vec{r},t)}{\partial
t{^2}}=0,
\end{equation}
For a free material particle, its potential energy $V(r)=0$ and
total energy $E=\frac{p^2}{2m}$. Substitute for the plane wave
equation (5) to Eq. (10), we have
\begin{equation}
(\frac{i}{\hbar}p)^2\psi(\vec{r})+CmE(\frac{i}{\hbar}E)^2\psi(\vec{r})=0,
\end{equation}
The constant $C$ is
\begin{equation}
C=-2/E,
\end{equation}
We can obtain the time-dependent wave equation of material
particle
\begin{equation}
\nabla^2
\psi(\vec{r},t)-\frac{2m(E-V(r))}{E^2}\frac{\partial^2\psi(\vec{r},t)}{\partial
t^{2}}=0,
\end{equation}
When the potential energy of the system is not a function of time,
the time-dependent wave equation (13) has variable separable
solutions of the form
\begin{equation}
\psi(\vec{r},t)=\psi(\vec{r})f(t)
\end{equation}
Substitute (14) into (13) and divide through by $\psi(\vec{r},t)$
\begin{equation}
\frac{1}{E-V(r)}\frac{1}{\psi(\vec{r})}\nabla^2 \psi(\vec{r})=
\frac{2m}{E^2}\frac{1}{f(t)}\frac{d^2f(t)}{dt^2}=C,
\end{equation}
$C$ is a separation constant independent of $\vec{r}$ and $t$. We
have two equations as follows:
\begin{eqnarray}
&&[\nabla^2+CV(r)]\psi(\vec{r})=CE\psi(\vec{r}),  \\&&
\frac{d^2f(t)}{dt^2}=\frac{2mC}{E^2}f(t),
\end{eqnarray}
Obviously, the Eq. (8) is the same as Eq. (16). Comparing their
coefficient, we have
\begin{equation}
C=-2m/{\hbar}^2,
\end{equation}
Substitute for $C$ to Eq. (16)-(17) and obtain
\begin{equation}
[-\frac{{\hbar}^2}{2m}\nabla^2+V(r)]\psi(\vec{r})=E\psi(\vec{r}),
\end{equation}
\begin{equation}
\frac{d^2f(t)}{dt^2}+\frac{4m^2}{E^2{\hbar}^2}f(t)=0,
\end{equation}
The solution of Eq. (20) is
\begin{equation}
f(t)=Ae^{i\frac{2m}{E\hbar}t}+Be^{-i\frac{2m}{E\hbar}t},
\end{equation}
The complete solution of Eq. (13) is
\begin{equation}
\psi(\vec{r},t)=\psi(\vec{r})(Ae^{i\frac{2m}{E\hbar}t}+Be^{-i\frac{2m}{E\hbar}t}).
\end{equation}
Obviously, Eq. (13) is different from Schr\"{o}dinger's
time-dependent wave equation \cite{s2}
\begin{equation}
[-\frac{{\hbar}^2}{2m}\nabla^2+V(r)]\psi(\vec{r},t)=i\hbar\frac{\partial\psi(\vec{r},t)}{\partial
t},
\end{equation}
The Schr\"{o}dinger's time-dependent wave equation (23) can be
obtained by making classical momenta $\vec{p}$ and energy $E$ into
operators
\begin{equation}
\vec{p}\rightarrow -i\hbar\nabla, E\rightarrow i\hbar
\frac{\partial}{\partial t}
\end{equation}
In Eq. (13), the momenta $\vec{p}$ can be become operator, i.e.,
$\vec{p}\rightarrow -i\hbar\nabla$, but the energy $E$ cannot be
become operator $i\hbar\frac{\partial}{\partial t}$, since there
isn't unity time $t$ in relativistic many-particle system. So, it
is reasonable that the $E$ is a number in Eq. (13).

In the following, we will give the relativistic wave equations of
material particle. Firstly, we should extend the Fermat's
principle to covariance from
\begin{equation}
\delta \int n ds=0,
\end{equation}
where the $ds$ is four-dimension differential interval and it is
Lorentz-invariant. It is important to find a relativistic
variational equation of material particle. We can do it from
Hamilton's principle \cite{s4}
\begin{equation}
\delta \int L dt=0,
\end{equation}
where $L$ is Lagrange function. The Eq. (26) can be written by
$dt=\gamma d\tau$ and $ds=c d\tau$ as
\begin{equation}
\delta \int \frac{\gamma}{c}L ds=0,
\end{equation}
where $\gamma=\frac{1}{\sqrt{1-\frac{u^2}{c^2}}}$, the $u$ is
particle velocity, the $c$ is light velocity. Obviously, the Eq.
(27) is covariance when the $\gamma L$ is Lorentz-invariant. For a
free particle, the $L$ can be taken as \cite{s5}:
\begin{equation}
L=-m_0c^2\sqrt{1-\frac{u^2}{c^2}},
\end{equation}
From Eqs. (25)-(28), with the same method in the above, we can
obtain the time-independent relativistic wave equations of
material particle
\begin{equation}
[E_0^2-c^2{\hbar}^2{\nabla}^2]\psi(\vec{r})=E^2\psi(\vec{r}),
\end{equation}
where $E_0=m_0c^2$. The equation of classical mass-energy is
\begin{equation}
E^2=E_0^2+c^2p^2,
\end{equation}
Obviously, the Eq. (29) also can be obtained by quantization to
Eq. (30), i.e., $\vec{p}\rightarrow -i\hbar \nabla$. However, the
time-dependent relativistic wave equations of material particle
can be obtained easily. It is obtained similarly to Eq. (13)
\begin{equation}
\nabla^2
\psi(\vec{r},t)-\frac{E^2-E_0^2}{E^2c^2}\frac{\partial^2\psi(\vec{r},t)}{\partial
t^{2}}=0,
\end{equation}
where the $E$ is total energy. The Eq. (31) is different from the
Klein-Gordon equation
\begin{equation}
\nabla^2 \psi(\vec{r},t)-\frac{m_0^2c^2}{{\hbar}^2}\psi(\vec{r},t)
-\frac{1}{c^2}\frac{\partial^2\psi(\vec{r},t)}{\partial t^{2}}=0,
\end{equation}
and the Eq. (31) has variable separable solutions of the form
\begin{equation}
\psi(\vec{r},t)=\psi(\vec{r})f(t),
\end{equation}
substituting Eq. (33) into (31), we can obtain two equations as
follows:
\begin{equation}
[E_0^2-c^2{\hbar}^2{\nabla}^2]\psi(\vec{r})=E^2\psi(\vec{r}),
\end{equation}
\begin{equation}
\frac{d^2f(t)}{dt^2}+\frac{E^2}{{\hbar}^2}f(t)=0,
\end{equation}
The Eq. (34) is the same as (29). The complete solution of Eq.
(31) is
\begin{equation}
\psi(\vec{r},t)=(Ae^{i\frac{E}{\hbar}t}+Be^{-i\frac{E}{\hbar}t})\psi(\vec{r}).
\end{equation}

In the following, we consider a charged particle in an
electromagnetic field. In Eq. (27), the $L$ can be taken as
\cite{s5}:
\begin{equation}
L=-\frac{m_0c^2}{\gamma}-\frac{e}{\gamma c}U_{\mu}A^{\mu}
=-\frac{m_0c^2}{\gamma}+\frac{e}{c}\vec{u}\cdot\vec{A}-e\Phi,
\end{equation}
where the $U_{\mu}$ is four-velocity, the $A^{\mu}$ is
electromagnetic four-vector. The momentum $\vec{P}$ conjugate to
$\vec{X}$ can be defined as follows:
\begin{equation}
P_i=\frac{\partial L}{\partial u_i}=\gamma mu_i+\frac{e}{c}A_i,
\end{equation}
or
\begin{equation}
\vec{P}=\vec{p}+\frac{e}{c}\vec{A}=\gamma
m_0\vec{u}+\frac{e}{c}\vec{A},
\end{equation}
where $\vec{p}=\gamma m\vec{u}$. From Eq. (39), we have
\begin{equation}
\vec{u}=\frac{c\vec{P}-e\vec{A}}{\sqrt{(\vec{P}-\frac{e}{c}\vec{A})^2+m_0^2c^2}},
\end{equation}
\begin{equation}
\gamma=\frac{\sqrt{(\vec{P}-\frac{e}{c}\vec{A})^2+m_0^2c^2}}{m_0c},
\end{equation}
The total energy $E$ is
\begin{equation}
E=\sqrt{(c\vec{P}-e\vec{A})^2+m_0^2c^4}+e\Phi,
\end{equation}
From Eqs. (37)-(42), we obtain
\begin{equation}
\gamma L=-m_0c^2+\gamma \frac{e}{c}\vec{u}\cdot\vec{A}-\gamma
e\Phi=-m_0c^2+\frac{ec}{E_0}\vec{p}\cdot\vec{A}-\frac{E-e\Phi}{E_0}e\Phi,
\end{equation}
with the same method in the above, we can obtain the
time-independent wave equation when $n$ is replaced with $\gamma
L$ in Eq. (1). It is
\begin{equation}
\nabla^2\psi(\vec{r})+B(-E_0+\frac{ec}{E_0}\vec{p}\cdot\vec{A}-\frac{E-e\Phi}{E_0}e\Phi)^2\psi(\vec{r})=0,
\end{equation}
for a free particle
\begin{equation}
\vec{A}=\Phi=0,
\end{equation}
and
\begin{equation}
\psi(\vec{r})=\frac{1}{(2\pi \hbar)^{3/2}}
e^{\frac{i}{\hbar}(\vec{p}\cdot\vec{r}-Et)},
\end{equation}
substitution of Eqs. (45)-(46) into (44) gives
\begin{equation}
B=\frac{E^2-E_0^2}{E_0^2c^2{\hbar}^2},
\end{equation}
where $E=\gamma m_0c^2$ for a free particle. Substitute for
$\gamma$ to obtain
\begin{equation}
B=\frac{(c\vec{P}-e\vec{A})^2}{E_0^2c^2{\hbar}^2}=\frac{{p}^2}{E_0^2{\hbar}^2},
\end{equation}
\begin{equation}
{\nabla}^2\psi(\vec{r})+\frac{{p}^2}{E_0^2{\hbar}^2}(-E_0+\frac{ec}{E_0}\vec{p}\cdot\vec{A}
-\frac{E-e\Phi}{E_0}e\Phi)^2\psi(\vec{r})=0,
\end{equation}
Substitute for $\vec{p}\rightarrow -i\hbar \nabla$ to obtain
\begin{equation}
{\nabla}^2\psi(\vec{r})-\frac{1}{E_0^2}{\nabla}^2(-E_0-i\hbar\frac{ec}{E_0}\nabla\cdot\vec{A}-
\frac{E-e\Phi}{E_0}e\Phi)^2\psi(\vec{r})=0.
\end{equation}
The Eq. (50) is time-independent relativistic wave equation of a
charged particle in an electromagnetic field. For time-dependent
relativistic wave equation, we can obtain when $n$ is replaced
with $\gamma L$ in Eq. (9). It is
\begin{equation}
{\nabla}^2\psi(\vec{r},t)+B(-E_0+\frac{ec}{E_0}\vec{p}\cdot\vec{A}-\frac{E-e\Phi}{E_0}e\Phi)^2
\frac{\partial^2\psi(\vec{r},t)}{\partial^2 t}=0,
\end{equation}
We can obtain the constant $B$ Similarly to Eq. (48)
\begin{equation}
B=-\frac{p^2}{E_0^2(E-e\Phi)^2},
\end{equation}
substitution of Eq. (52) into (51) gives
\begin{equation}
{\nabla}^2\psi(\vec{r},t)-\frac{p^2}{E_0^2(E-e\Phi)^2}(-E_0+\frac{ec}{E_0}\vec{p}\cdot\vec{A}
-\frac{E-e\Phi}{E_0}e\Phi)^2\frac{\partial^2\psi(\vec{r},t)}{\partial^2
t}=0,
\end{equation}
Substitute for $\vec{p}\rightarrow -i\hbar \nabla$ to obtain
\begin{equation}
{\nabla}^2\psi(\vec{r},t)+\frac{{\hbar}^2}{E_0^2(E-e\Phi)^2}{\nabla}^2(-E_0-i\hbar\frac{ec}{E_0}\nabla\cdot\vec{A}-
\frac{E-e\Phi}{E_0}e\Phi)^2\frac{\partial^2\psi(\vec{r},t)}{\partial^2
t}=0.
\end{equation}
The Eq. (54) is time-dependent relativistic wave equation of a
charged particle in an electromagnetic field.

In our works, the time-independent wave equations are the same as
the standard quantum mechanics. So, the energy of system is same
in the two theoretical calculation. Since our time-dependent wave
equations are different from standard quantum mechanics the
systematic transition probability is different in their
calculation. Otherwise, the new time-dependent relativistic wave
equations may have some effects to quantum field theory.

\end{document}